\documentclass{optica-article}

\journal{opticajournal} % for journals or Optica Open

\articletype{Research Article}

\usepackage{lineno}
\newcommand{\e}{\mathrm{e}}
\newcommand{\Sym}{\mathcal{S}}
\newcommand{\uniform}{\mathcal{U}}
\newcommand{\tildet}{\Tilde{t}}
\newcommand{\matx}{\mathbf{x}}
\newcommand{\maty}{\mathbf{y}}
\newcommand{\ybar}{\bar{y}}
\newcommand{\matytar}{\bar{\mathbf{y}}}
\newcommand{\matWRO}{\mathbf{W}_\mathrm{RO}}
\newcommand{\ahat}{\hat{a}}
\newcommand{\SNReta}{\mathrm{SNR}_\eta}
\newcommand{\SNRtheta}{\mathrm{SNR}_\vartheta}

\newcommand{\revision}[1]{#1}

\begin{document}

\title{Photonic reservoir computing enabled by stimulated Brillouin scattering}

\author{Sendy Phang}

\address{George Green Institute for Electromagnetics Research, Faculty of Engineering, University of Nottingham, University Park, NG7 2RD, UK}

\email{\authormark{*}sendy.phang@nottingham.ac.uk} %% email address is required; see note below about the corresponding author designation

% use {asbstract*} to suppress the copyright line. Copyright information will be added in production

\begin{abstract*} 
Artificial Intelligence (AI) drives the creation of future technologies that disrupt the way humans live and work, creating new solutions that change the way we approach tasks and activities, but it requires a lot of data processing, large amounts of data transfer, and computing speed. It has led to a growing interest of research in developing a new type of computing platform which is inspired by the architecture of the brain specifically those that exploit the benefits offered by photonic technologies, fast, low-power, and larger bandwidth. Here, a new computing platform based on the photonic reservoir computing architecture exploiting the non-linear wave-optical dynamics of the stimulated Brillouin scattering is reported. The kernel of the new photonic reservoir computing system is constructed of an entirely passive optical system. Moreover, it is readily suited for use in conjunction with high performance optical multiplexing techniques to enable real-time artificial intelligence. Here, a methodology to optimise the operational condition of the new photonic reservoir computing is described which is found to be strongly dependent on the dynamics of the stimulated Brillouin scattering system. The new architecture described here offers a new way of realising AI-hardware which highlight the application of photonics for AI.

\end{abstract*}

%%%%%%%%%%%%%%%%%%%%%%%%%%  body  %%%%%%%%%%%%%%%%%%%%%%%%%%
\section{Introduction}
Photonic reservoir computing (PhRC) is a physical realisation of an artificial neural network (ANN) architecture based on reservoir computing (RC) as a photonic system \cite{larger2012photonic,paquot2012optoelectronic,duport2012all,phang2019optical}. PhRC mimics the way the nervous system process information in a distributed manner, in contrast to the centralised approach employed in a digital central processing unit (CPU) which is based on a von Neumann architecture \cite{shastri2017principles}. The central motivation in developing a neuromorphic computing system is to have a closer integration between the hardware and information processing/computing, enabling a hardware-level integration of artificial intelligence. Neuromorphic computing systems other than PhRC are also actively explored, including photonic synapses \cite{cheng2017chip,zhang2020recent}, photonic spike processors \cite{robertson2022ultrafast}, and photonic neural networks \cite{feldmann2021parallel}. Neuromorphic photonic systems are designed to be inherently suited for AI operations and promise faster ($10^6$-fold more calculations per second per $\mathrm{m}^2$ surface-area) and more energy-efficient ($10^3$-fold more calculations per second per Watt) operations compared to traditional micro-electronics processors \cite{shastri2017principles}, making them a highly promising choice for the development of a new photonic AI processor. 

The RC scheme is the underlying concept of the new PhRC proposed in this present work; readers are referred to \cite{jaeger2004harnessing,lukovsevivcius2012reservoir,appeltant2011information} for a comprehensive review on the reservoir computing scheme. The RC is a relatively new high-dimensional ANN computing scheme with \revision{recurrent neuron interconnections}; thus, it is a variant of the recurrent neural network \cite{jaeger2004harnessing,lukovsevivcius2012reservoir}. Being part of the recurrent neural network family, the RC is suited for a time-series based task. The RC scheme consist of two main parts, namely a kernel and read-outs. The kernel is comprised of randomly connected non-linear neurons with recurrent pathways; as such the neuron activation state of each neuron in the kernel is a systematic variant of the driving time-dependent input signal. The read-out layers perform a final inference based on the sampled neuron activation states of the neurons in the kernel. The PhRC’s operating principle is fundamentally different compared to other neuromorphic approaches. The PhRC kernel is a semi-chaotic high-order dynamical photonic system that maps input signals into a higher-dimensional feature space through its dynamic behaviour. A linear estimator is applied at the read-out to infer potential relationships/patterns in the transformed input data. For the PhRC, only the read-out undergoes objective-specific training. This approach is made possible by the kernel's ability to transform the input signal into a higher dimensional representation, which can be analysed using a simple linear estimator at the Read-out \cite{jaeger2004harnessing,duport2012all,appeltant2011information,paquot2012optoelectronic,larger2012photonic}. The PhRC system offers practical advantages over other neuromorphic systems, viz.: (i) it offers different ways of realisation of higher-dimensional photonic hardware system, e.g., using splitters, cavities, laser, amplifiers, etc \cite{van2017advances}; (ii) the PhRC’s read-out is designed by a fast, globally optimal linear regression while other neuromorphic systems are trained by a slow-converging backpropagation or neural engineering framework \cite{shastri2017principles}; (iii) the PhRC utilises optical multiplexing to create highly scalable nodes, whereas other approaches require physical nodes to be created\cite{shastri2017principles}. Several implementations of RC as a photonic system, PhRC, have been reported, for example as a time-delayed fibre loop architecture with electro-optical components \cite{paquot2012optoelectronic}, or as a fully optical system exploiting non-linear characteristic of a semi-conductor amplifier \cite{vandoorne2011parallel}, or as a photonic crystal cavity \cite{laporte2018numerical,phang2021neuromorphic,phang2019optical}. In this paper, a new architecture of PhRC, which is based on an optical-fibre and loop-free kernel configuration is reported. A non-linear neuron activation function is achieved by means of a non-linear stimulated Brillouin photon-phonon scattering process.

Brillouin scattering is a well-known light-matter interaction phenomenon and has been extensively studied since the early 1920s \cite{brillouin1922diffusion}. The scattering phenomenon originates from the interaction of a photon from incident light with the medium, due to electrostriction, which modifies the material’s mechanical density profile generating a backscattered photon and a phonon. The scattered photon losses energy which is observed as a red shifting in frequency. This scattered photon is called a Stokes photon after George Stokes who first observed this red shifting phenomenon. The present work will consider Brillouin scattering phenomena occurring in a single-mode optical fibre. In such a scenario, the counter-propagating Stokes light interferes with the incident (pump) light which enhances the electrostriction process, creating a feedback loop, known as stimulated Brillouin scattering (SBS). The stimulated Brillouin scattering is a non-linear interaction between three-waves (pump, Stokes, and acoustic waves). Although SBS can be detrimental in optical telecommunication applications \cite{cotter1983stimulated}, it has been exploited and used in a range of applications, including radio-over-fibre, laser generation, sensors, etc. \cite{kobyakov2010stimulated,eggleton2019brillouin,merklein2017chip,zhu2007stored}. Related to the fundamental requirement of PhRC of memory, a SBS system has been shown to inherently possess memory property; applications as  optical memory devices have been demonstrated \cite{zhu2007stored,lecoeuche2000transient,merklein2017chip,eggleton2019brillouin}. Although in the present work in-fibre SBS phenomena are considered, it is noted that there is active research in developing a Brillouin scattering system in an integrated circuit, see Review papers \cite{eggleton2019brillouin}. It is believed that the work presented in this work will be relevant in the realisation of all optical PhRC based on SBS in a photonic chip.  

The following section details the architecture of the new PhRC system based on a stimulated Brillouin scattering system, i.e., PhRC-SBS. It describes the governing three-wave interaction equations used to model the PhRC-SBS system. It defines the optical multiplexing used to encode information and the demultiplexing approach.  Section 3 analyses the wave optical dynamics of the SBS system in relation to the performance of the PhRC-SBS system. Section 4 evaluates the performance of the PhRC-SBS to perform machine learning benchmark tests, including memory capacity evaluation, prediction of the non-linear dynamical system and for a communication channel equaliser. Finally, section \revision{5} summarises the findings reported in this paper. 

\section{Photonic reservoir computing with stimulated Brillouin scattering fibre kernel}

\begin{figure}[t!]
\centering\includegraphics[width=0.8\textwidth]{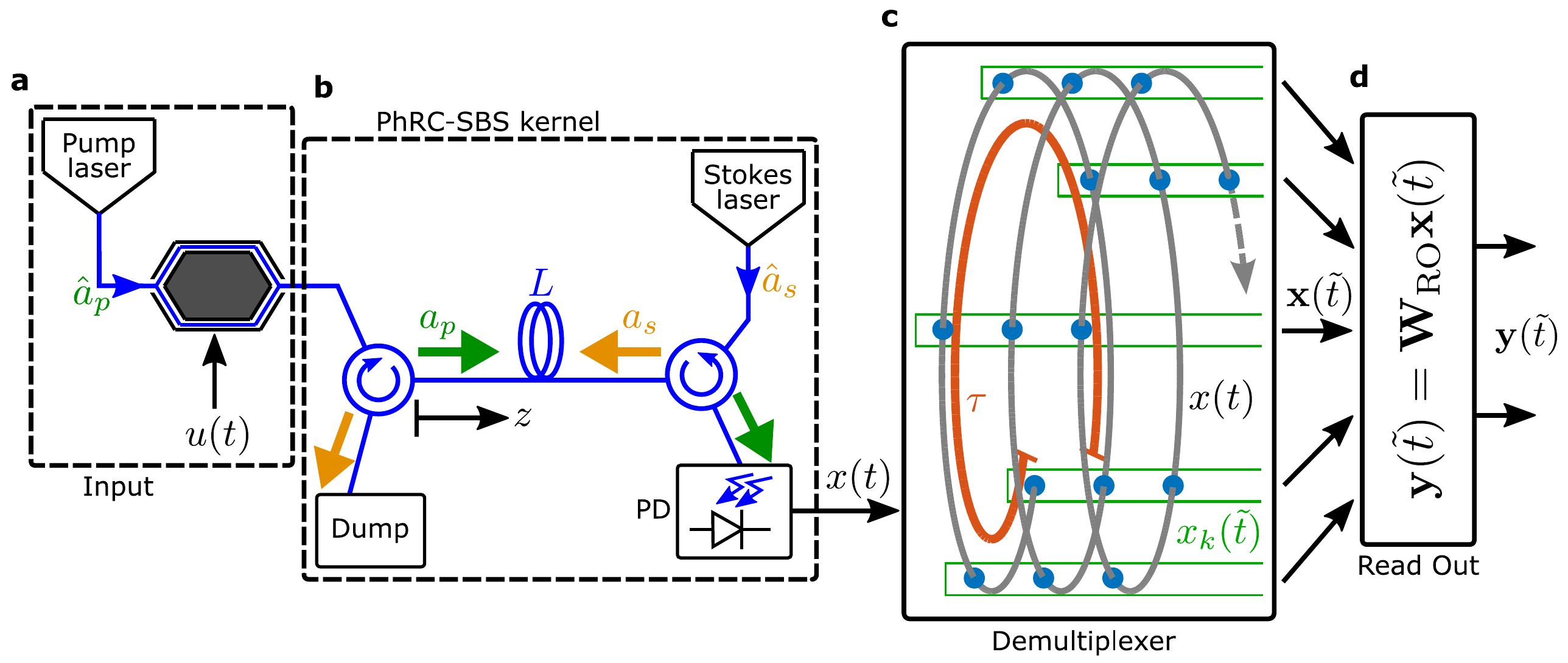}
\caption{Schematic of the PhRC-SBS system. (a) The input, (b) the kernel, (c) demultiplexer and (d) the linear read-out.}
\label{fig:phrcsbs}
\end{figure}

In the present work, the PhRC-SBS system is comprised of input, kernel, demultiplexer and read-out, as shown in Fig. \ref{fig:phrcsbs}.  It utilises two light sources, i.e., the pump and the Stokes continuous-wave sources. The Stokes source operates at frequency $\nu_s$ and the pump source operates at frequency $\nu_p$ with $\nu_p>\nu_s$.  Information is temporally multiplexed to the pump light using a Mach-Zehnder modulator via electrical biasing, which typically is done in practice by using an arbitrary waveform generator. The kernel is constructed of a spool of optical fibre in which the pump and Stokes laser signals propagate in opposing directions to stimulate Brillouin scattering. The circulators have been included at both of the fibre’s ends to avoid backscattered light from the photodetector and dumping port. As the pump and Stokes light only coexist in the kernel, the three-wave non-linear SBS interaction only occurs in the kernel part of the PhRC-SBS. The neuron activation state signal is monitored by the photodetector upon which it undergoes demultiplexing and is subsequently inferred by the read-out.  In the following subsection, the roles of each part of the PhRC-SBS are described.  

\subsection{Reservoir kernel based on stimulated Brillouin scattering}
In this paper, the three-wave interaction (TWI) coupled partial differential equations \cite{chow1993chaotic,lecoeuche2000transient,kobyakov2010stimulated} are used to model the SBS non-linear interaction which occurs in the kernel of the PhRC-SBS. The TWI equations are given by 
 \cite{chow1993chaotic,lecoeuche2000transient,kobyakov2010stimulated}, 
\begin{align} \label{eq:1}
    \frac{\partial}{\partial t} E_p + \frac{c}{n}\frac{\partial}{\partial z} E_p +\gamma_p E_p &= -K E_s \varrho \e^{-j\delta t} \notag\\
    \frac{\partial}{\partial t} E_s - \frac{c}{n}\frac{\partial}{\partial z} E_s +\gamma_s E_s &= K^* E_p \varrho^* \e^{j\delta t} \\
    \frac{\partial}{\partial t} \varrho + v_a \frac{\partial}{\partial z} \varrho + \gamma_a \varrho &= K^* E_p E_s^* \e^{j\delta t} \notag
\end{align}
where $E_p$, $E_s$ and $\varrho$ are the complex-valued slowly varying envelope of the pump, Stokes electric fields and acoustic wave, respectively; $c$ and $v_a$ are  \revision{the speed of light in vacuum and sound in the material}, \revision{respectively}; and $n$ is the effective index of the propagating waveguide mode. Attenuation of the pump, Stokes, and acoustic waves are given by $\gamma_p$, $\gamma_s$, and $\gamma_a$ parameters, respectively. The light and acoustic waves coupling, i.e., photon-phonon interaction, is given by the Brillouin coupling constant $K$. Parameter $\delta=2\pi(\nu_p-\nu_s-\delta \nu^\text{res})$ describes the resonance detuning condition, where $\nu_p$, $\nu_s$ and $\delta \nu^\text{res}$ are the pump, Stokes, and Brillouin shift frequency, respectively. As such, the interaction is at resonance when $\nu_p-\nu_s=\delta\nu^\text{res}$ and at off-resonance when $\nu_p-\nu_s\ne\delta\nu^\text{res}$. The resonance detuning $\delta$ is amendable  via chemical doping or mechanical stress due to temperature change or strain along the fibre; this dependence has been exploited for application in distributed sensor systems \cite{lecoeuche2000transient,kobyakov2010stimulated}. 

The TWI equations (\ref{eq:1}) can be simplified by taking normalisations of variables $T=t\gamma_a$, $Z=z\gamma_a (n/c)$, $\Delta=\delta/\gamma_a$, and variable substitutions $a_p\rightarrow E_p K/\gamma_a$,  $a_s\rightarrow E_s K/\gamma_a$, and $a_a\rightarrow\varrho K\e^{j\delta t}/\gamma_a$.  Then, further considering the case of that modern telecom-grade optical fibre has a very low loss, e.g., attenuation of standard SMF-28 $\gamma_{p,s}<0.18$ dB/km \cite{corningsmf28}, it yields to the situation whereby the photon-phonon interaction occurs on a timescale much faster than the propagation of the acoustic wave, as such (\ref{eq:1}) can be expressed as \cite{chow1993chaotic,lecoeuche2000transient}
\begin{align} \label{eq:2}
    \frac{\partial}{\partial T} a_p + \frac{\partial}{\partial Z} a_p &= -a_s a_a  \notag\\
    \frac{\partial}{\partial T} a_s - \frac{\partial}{\partial Z} a_s &= a_p a_a^* \\
    \frac{\partial}{\partial T} a_a + (1+j\Delta) a_a  &= a_p a_s^* \notag
\end{align}

Throughout this work, the normalised TWI equations, (\ref{eq:2}), are solved numerically using the finite-difference approach \cite{marble2004stimulated}. The present work considers SBS in a standard SMF-28 fibre as in \cite{lecoeuche2000transient} which uses Brillouin coupling parameters $K=80 \text{ ms}^{-1}\mathrm{V}^{-1}$ and $\gamma_a=200$ MHz (corresponding to 35 MHz resonance-linewidth). The kernel, a standard SMF-28 fibre, is operating around the 1.3~$\mu$m wavelength with an effective mode area $A=85\text{ }\mu \mathrm{m}^2$ and effective index $n=1.467$\cite{corningsmf28}. Using these realistic parameters: an optical fibre length in the normalised unit $L=1$ corresponds to a physical length of 1.02 m, normalised detuning parameter $\Delta=1$ corresponds to $\delta=200$ MHz, and $a_{p,s}=1$ corresponds to $E_{p,s}=2.5$ $\mathrm{MVm}^{-1}$ from which the optical power $P$ can be calculated from the normalised pump or Stokes amplitude $a_{p,s}$ via $P =AnE_{p,s}^2/(2Z_0)\approx1.034a_{p,s}^2 \text{ (in Watt)}$ with $Z_0$ the free-space wave impedance. For generality, normalised variables are used from hereon.

\subsection{PhRC-SBS input} \label{sec:in}

In the present work, let the Stokes light be a continuous wave propagating from the right to left and not carrying any information, see Fig. \ref{fig:phrcsbs}. The information symbol stream is encoded onto the pump light which is injected from the left-end side of the fibre. It will be seen, in Section \ref{sec:benchmark}, that different types of input symbol will be used depending on the task to be solved. An optical multiplexing technique is utilised to encode the information symbol, $\mathcal{S}$, as a temporally amplitude modulated signal.

The encoding scheme is described as follows: first, the raw symbol $\Sym$ is normalised as such its entries are valued between $\chi$ and 1 via $\bar{\Sym}=(1-\chi)\frac{\Sym-\min[\Sym]}{\max[\Sym] - \min[\Sym]} + \chi$, where $0<\chi<1$. Then, a sample and hold operation is performed, as such the normalised symbols $\bar{\Sym}$ are serialised in time, each with a duration of $\tau$ to yield the temporal signal $J$. Next, a modulation is applied to $J$ by a periodic mask signal, $m(t)=m(t+\tau)$, where $m\sim(1+\uniform(-\hat{m},\hat{m}))$. Here, $\uniform(-\hat{m},\hat{m})$ denotes a uniformly distributed random number generator over a limited interval of $(-\hat{m},\hat{m})$. This way, the mask signal fluctuates around $\hat{m}$ with a mean value of 1. The mask signal is piecewise constant over a period of $\theta=\tau/N_x$, where $N_x$ denotes the number of masks per mask period $\tau$. Finally, the masked information signal is adjusted by the information modulation depth $\hat{\kappa}$. Concisely, the modulated pump signal $a_p$  is given by, 
\begin{align}\label{eq:3}
a_p(t)=\ahat_p u(t),    
\end{align}
where
\begin{align}
u(t)=1+\hat{\kappa} m(t)J(t).    
\end{align}
In (\ref{eq:3}), $\hat{a}_p$ is the amplitude of the injected pump. This encoding scheme allows differentiation between the case of no-symbol and min-valued symbol; as such, in the case of no-symbol $\bar{\Sym}=\{\varnothing\}$, $a_p=\hat{a}_p$ and in the case of $\bar{\Sym}=\chi$, $a_p\approx\hat{a}_p (1+\hat{\kappa}\chi)$. In the case of no-symbol, the PhRC-SBS is referred to as being in the \textit{idle} state (operating but not processing any information).  Note that the encoding scheme described follows an approach used in\cite{larger2012photonic,paquot2012optoelectronic,appeltant2011information,anufriev2022non} in which the number of masks $N_x$ directly corresponds with the number of `virtual' neuron nodes. Other multiplexing techniques, e.g., phase modulation\cite{bauwens2022influence}, can be used. 

Although, in principle, the floor bound of $\bar{\Sym}$ can be of any value $0<\chi<1$, practically one needs to consider the resolution of the digital-to-analogue signal converter; throughout the present work $\chi=0.5$ has been used. Moreover, such a modulation is typically achieved by using an electro-optics Mach-Zehnder modulator\cite{anufriev2022non}. Linear operation of the electro-optical modulator is considered, here, meaning that it operates at the quad point and $\max[u] - \min [u] \ll V_\pi$, where $V_\pi$ denotes the half-wave voltage, i.e., the voltage needed to induce a $\pi$ phase shift between the two optical arms of the interferometer. It is emphasised here that the role of the modulator is only to encode the information stream to the pump laser and the PhRC-SBS does not exploit the non-linear property of the modulator for computing, in contrast to other PhRC systems \cite{larger2012photonic,paquot2012optoelectronic,anufriev2022non}. For definiteness, in the present work the following encoding parameters $\hat{\kappa}=\hat{m}=10\%$ are fixed throughout. To demonstrate the encoding schemed described, here, an illustrative example with a detailed process is provided in the Appendix (Subsection \ref{app:mul}).

\subsection{Demultiplexing and the read-out}
In the PhRC-SBS setup, the output of the kernel is monitored by the photodetector. Following the convention in the photonic reservoir computing community, the term of activation state signal $x(t)$ for the monitored output signal coming out from the kernel is used. As the information stream has been temporally multiplexed and masked at the input, it is necessary for the activation state signal to be demultiplexed and repartitioned in terms of its mask before being interpreted by the read-out. The present work has adopted a demultiplexing procedure which has been used previously \cite{larger2012photonic,paquot2012optoelectronic,anufriev2022non} which is briefly described as follows: first, the activation state signal is defined as the absolute value of the laser amplitude at the right-end of the fibre spool with an addition of noise, viz., $x(t)=|a_p|(t,L)+\eta$ where $\eta$ is an independent identically distributed (i.i.d) Gaussian noise with zero mean adjusted in power corresponding to a signal-to-noise ratio ($\mathrm{SNR}_\eta$), emulating the overall system’s noise. To demultiplex activation signal $x(t)$ in terms of each mask index, $x(t)$ is spun with a winding period of $\tau$. This scheme produces a discretised and partitioned activation state signal $x(\Tilde{t})=[x_1 (\Tilde{t}),\cdots,x_k(\Tilde{t})]^T$, where $k=1,2,\cdots,N_x$ denoting the index of the virtual nodes and each entry of matrix $\matx(\Tilde{t})$ are given by $x_k(\Tilde{t})\leftarrow x(\Tilde{t} \tau+k\theta)$ with $\Tilde{t}=0,1,\cdots$ being the winding index. Figure \ref{fig:phrcsbs}(c) illustrates the demultiplexing scheme. Alternatively, for a real-time application one can utilise an ultrafast all optical time-division demultiplexing technique\cite{kawanishi1998ultrahigh,gutierrez2001novel}.

The process subsequently following the demultiplexing is estimation. The present work utilises a linear estimation procedure in which the estimated output signal $y$ is constructed from the demuxed activation state matrix $\matx(\Tilde{t})$,
\begin{align}
    \maty(\tildet) = \matWRO \matx(\tildet), \quad \maty \in \mathbb{R}^{N_y\times 1}
\end{align}
where $\matWRO \in \mathbb{R}^{N_y\times N_x}$ is the read-out weighting and $N_y$ is the dimension of the output signal. For clarity, throughout this paper, the notation $\maty$ and $\matytar$ is used to denote the estimated and target output signals, respectively. The read-out weight is obtained from a training session by a Tikhonov regularisation with cross-validation as in \cite{lukovsevivcius2012reservoir,phang2019optical}. There are other estimation schemes, including those which use constant bias and/or auto-regression as in \cite{lukovsevivcius2012reservoir} or an online Kalman filter \cite{jin2022adaptive}. Here, the linear estimation scheme is used as it is the simplest estimator and allows comparison with other published PhRC configurations which have used the same linear estimation approach. It is further assumed, here, that inclusion of constant bias and/or auto-regression will yield to better estimation. 

\section{Dynamics of PhRC kernel based on stimulate Brillouin scattering} \label{sec:dynamics}

This section analyses the dynamics of the PhRC-SBS kernel whose underlying non-linear property is originated from the stimulated Brillouin scattering (SBS). It will show that the dynamics of the SBS undergoes a phase transition from a \revision{single-} to a \revision{multi-states} wave-\revision{scattering system} which in the context of the present work directly influences its performance as a PhRC system. The SBS phenomena have  been studied extensively analytically and experimentally, see \cite{chow1993chaotic,kobyakov2010stimulated,eggleton2019brillouin}. Its dynamics are known to exhibit phase transition from \revision{single-} to \revision{multi-states} behaviour. There are a few tuneable parameters in (\ref{eq:2}) which define is overall dynamics, including, the amplitude of the driving pump and Stokes light $\hat{a}_{p,s}$ and the detuning parameter $\Delta$. Moreover, it is expected that the total optical fibre $L$ which defined the non-linear interaction length will also have impact on the phase transition. To analyse the dynamics of the SBS, in this section the PhRC-SBS is operating in its idle state. 

\begin{figure}[b!]
\centering\includegraphics[width=0.8\textwidth]{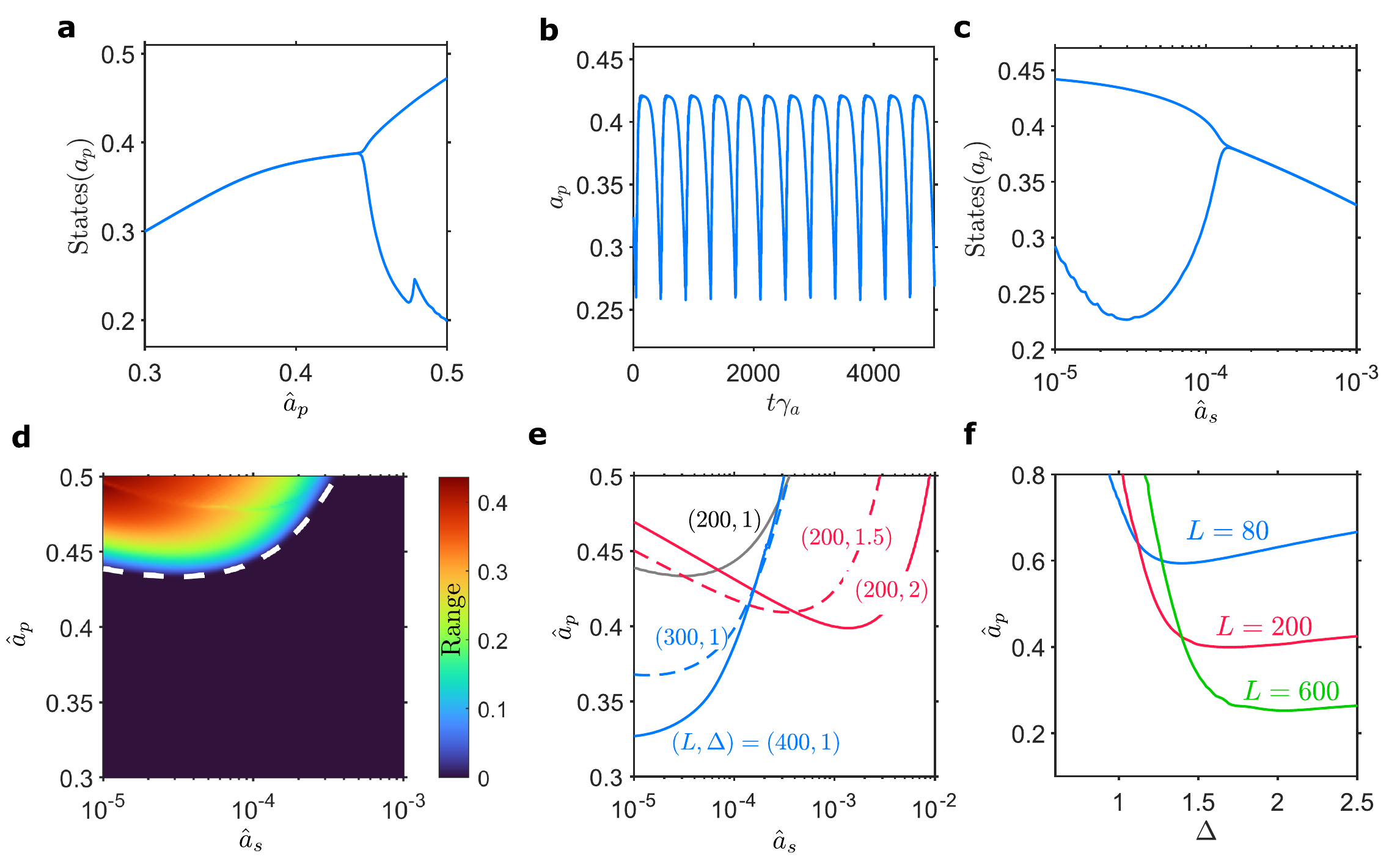}
\caption{Phase transition from \revision{single-} to \revision{multi-states} of the SBS system. (a) Min and max states of the pump light as a function of $\ahat_p$ with $\ahat_s=10^{-4}$. (b) The temporal signal $a_p(L)$ for $\ahat_p=0.46$, in the \revision{multi-states} regime with $\ahat_s=10^{-4}$. (c) Min and max state of the pump light as a function of $\ahat_s$ with $\ahat_p=0.45$. (d) The range of pump light states in the $(\ahat_p,\ahat_s)$ parameters space. (e) Impact of $L$ and $\Delta$ to the bifurcation line. (f) Bifurcation lines of the SBS system for various $L$ in the $(\Delta,\ahat_p)$ parameter space for a fixed value of $\ahat_s=2.5\times10^{-3}$.}
\label{fig:dynamic}
\end{figure}

Figure \ref{fig:dynamic}(a) shows the maximum and minimum states of the pump laser at the end of the fibre spool $a_p (L)$ as a function of the injected pump $\ahat_p$ for a fixed driving Stokes laser with amplitude $\ahat_s=10^{-4}$ at the steady state. It shows that at the steady states, $a_p(L)$ is single valued, i.e., $\max[a_p(L)]=\min[a_p(L)]$, until $\ahat_p\approx 0.44$ from which point it splits into two states. This point is referred to as the bifurcation point, which denotes the transition between \revision{single-} and \revision{multi-states} solutions. This bifurcation is exemplified in Fig. \ref{fig:dynamic}(b) which depicts the steady state of $a_p (t,L)$ operating at $\ahat_p=0.46$ after the bifurcation point. Phase transition behaviour also can be observed by varying the driving Stokes laser amplitude $\ahat_s$. Figure \ref{fig:dynamic}(c) shows the $\max[a_p(L)]$ and $\min[a_p(L)]$ as a function of $\ahat_s$ for a fixed driving pump $\ahat_p=0.45$. It shows the presence of a coalescence point, i.e., $a_p (L)$ is double valued at low value of $\ahat_s$ and coalesce to a single valued state at this point. To provide an overall picture of the dynamics, Fig. \ref{fig:dynamic}(d) depicts the range of $a_p (L)$, i.e., $\max[a_p(L)]-\min[a_p(L)]$, for both the $\ahat_p$ and $\ahat_s$. It indicates two distinct regions where $a_p (L)$ is either \revision{single-state} or not, which in general the \revision{multi-states} region occurs at low $\ahat_s$  and high $\ahat_p$ values. The dashed line in Fig. \ref{fig:dynamic}(d) indicates the phase transition of the range manifold which is referred to as the bifurcation line from hereon. 

As was mentioned, the dynamics of the SBS system is expected to be influenced also by the physical length of the fibre  $L$ and the detuning parameter $\Delta$. Figure \ref{fig:dynamic}(e) depicts the impact of varying $L$ and $\Delta$ to the phase transition of the SBS system. It shows that by increasing the length of fibre $L$, phase transition happens at lower driving pump amplitude while increasing the detuning parameter $\Delta$ shifts the overall bifurcation line to a higher injected Stokes amplitude. Figure \ref{fig:dynamic}(\revision{f}) shows that the bifurcation lines resemble the well-known L-shaped bifurcation lines in the $(\Delta,\ahat_p)$ parameter space as previously reported in \cite{chow1993chaotic}. It also affirms that increasing the length reduces the required $\ahat_p$ to reach phase transition.

\section{Benchmark} \label{sec:benchmark}
Optimum operation for a neuromorphic PhRC system based on a time-delayed architecture has been shown previously to be around the phase transition of the kernel\cite{larger2013complexity,anufriev2022non}. It can be rationalised that at the bifurcation point the dynamics of the kernel is highly non-linear and sensitive to perturbation, which are key requirements for reservoir computing\cite{paquot2012optoelectronic}.  In Section \ref{sec:dynamics}, the presence of the bifurcation line was observed; in this section the performance of the PhRC-SBS is evaluated and benchmarked to performing complex machine learning operations, including memory capacity estimation, prediction of a non-linear dynamical system, and for a communication channel equaliser. It will confirm that optimum operating conditions for these applications are indeed near the bifurcation line. 

\subsection{Memory capacity}

Systems exhibiting SBS are known to exhibit a memory property which has been exploited for a random-access memory device application\cite{zhu2007stored,merklein2017chip}. This memory property occurs because light propagates much faster than the acoustic wave, and the generated acoustic wave from the SBS retains some memory from this counter-propagating non-linear wave interaction. In this section, the (fading) memory property of the PhRC-SBS system is measured following the framework described in \cite{dambre2012information,duport2012all,appeltant2011information} which provides a quantitative measurable indicator in the context of neuromorphic computing. In this framework, memory is measured by calculating the total capacity to reconstruct a set of memory-dependent functions \cite{dambre2012information,duport2012all,appeltant2011information}. These functions' response depends on their  past value, i.e., they are recurrent, and can either be linearly or non-linearly dependent which yields linear and nonlinear memory capacities, respectively. The memory capacity $C$ can be calculated from the normalised mean squared error (NMSE) $\varepsilon$ of the reconstruction of the memory-dependent function $y$ via $C=1-\varepsilon[y]$. Other ways to calculate $C$ based on mutual- and self-correlation formulations can be found in \cite{dambre2012information}. Throughout this paper, NMSE is calculated by, 
\begin{align}
    \varepsilon = \frac{\langle (\maty - \matytar)^2 \rangle}{\langle (\matytar - \langle\matytar\rangle)^2 \rangle},
\end{align}
where $\langle\cdot \rangle$ denotes an ensemble averaging operation. For the linear fading memory, the PhRC-SBS is tasked to recall past input information symbol $\Sym$, i.e., 
\begin{align}
    \matytar_k^\mathrm{lin} (\tildet) = \Sym (\tildet-k), \quad k = 1,2,\cdots  ,
\end{align}
where the input symbol is $\Sym\sim\mathcal{U}(-1,1)$. The total linear memory capacity is calculated by,
\begin{align} \label{eq:8}
    C_\mathrm{linear} = \sum_k C_k = \sum_k\big(1-\varepsilon[y_k]\big) .
\end{align}

Jaeger, \textit{et al}. in \cite{jaeger2004harnessing} showed that the total linear memory capacity (\ref{eq:8}) is less than the number of the computing kernel nodes, i.e., $C_\mathrm{linear}<N_x$. In \cite{dambre2012information}, Dambre, \textit{et al}. generalised this relation, showing that the overall capacity of constructing a memory-dependent function covering all possible orthogonal functions in the Hilbert space, $C_{H^\infty}$, is bounded by $N_x$, i.e., $C_{H^\infty}\leq N_x$; that is a PhRC with a perfect memory will have $C_{H^\infty}=N_x$. For orthogonal functions in the Hilbert space, one can use the whole set of Legendre or Chebyshev polynomials as the reconstruction target function. To allow comparison with other published memory capacities, the present work will only consider the second order Legendre polynomial reconstruction target function,
\begin{align}
    \matytar_k^\mathrm{quad} (\tildet) = 3\Sym^2 (\tildet-k), \quad k = 1,2,\cdots , 
\end{align}
and the product of two past values of the input information sequence symbol,
\begin{align}
    \matytar_{kk'}^\mathrm{cross} (\tildet) = \Sym (\tildet-k)\Sym (\tildet-k'), \quad k,k' = 1,2,\cdots  .
\end{align}

The total capacities for these non-linear memory reconstruction tasks, $C_\mathrm{quad}$ and $C_\mathrm{cross}$, are calculated in the same way as $C_\mathrm{linear}$. The impact of the pump’s and Stokes’ amplitude $\ahat_{p,s}$ and modulation time parameter $\tau$ to the memory capacities is now analysed. Throughout this analysis, the kernel’s optical fibre length $L=200$, and the SBS detuning parameter $\Delta=1$ are fixed. Also, the analysis uses 3000 symbols of which the first 80\% is used for training and the remaining for testing. For comparison purposes, the same number of masks $N_x=50$ and $k=100$ for the memory reconstruction tasks have been used; these are the same parameters as used in \cite{paquot2012optoelectronic,duport2012all}.

\begin{figure*}[tbp]
\centering
\includegraphics[width=0.99\textwidth]{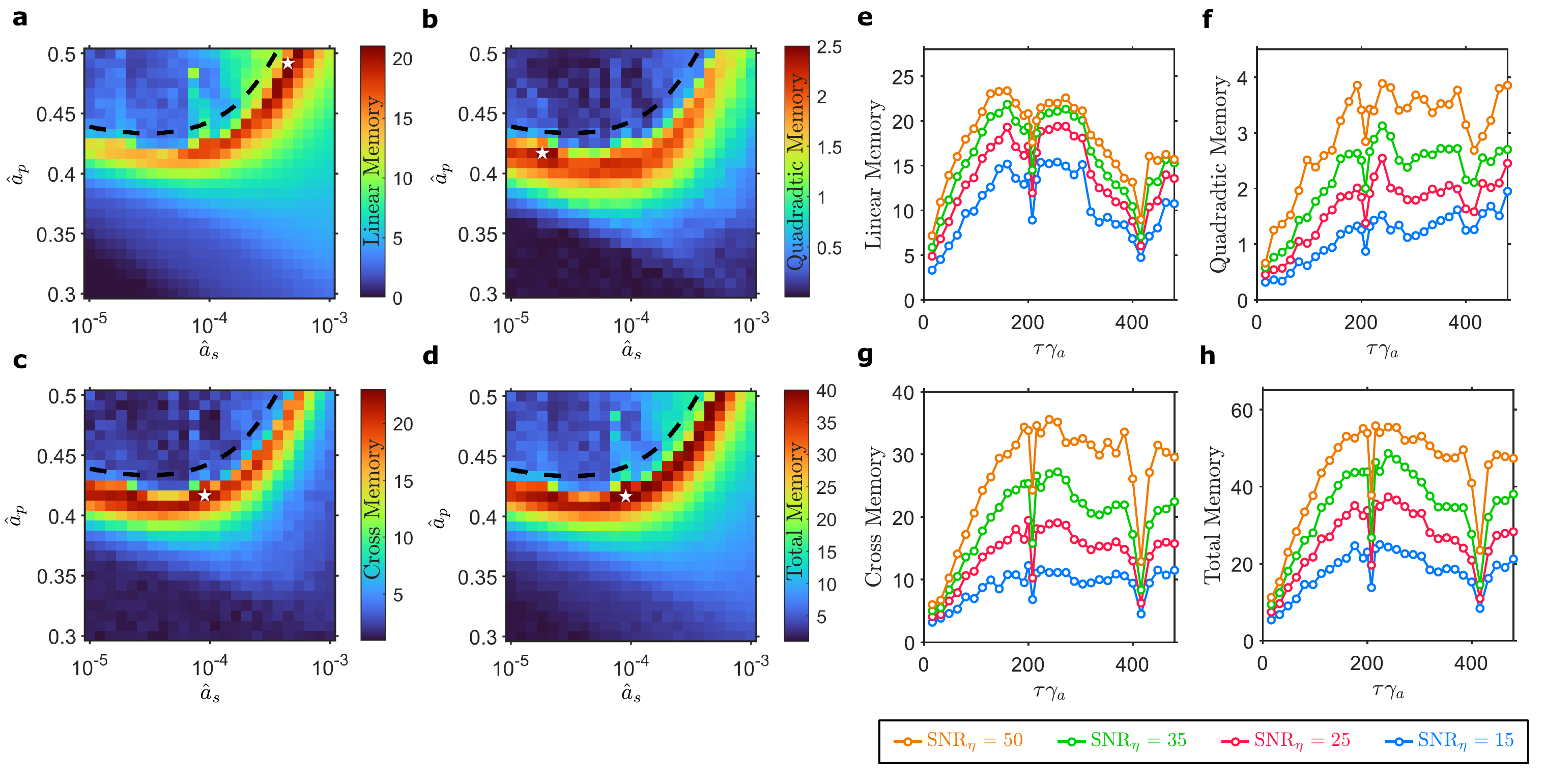}
\caption{Operational condition for PhRC-SBS to perform memory reconstruction tasks. (a-d) Memory capacities $C_\mathrm{linear}$, $C_\mathrm{quad}$, $C_\mathrm{cross}$ and $C_\mathrm{total}$ as function of pump and Stokes amplitude $\ahat_{p,s}$. The maximum capacity found within the parameter space has been marked by a star. (e-h) The maximum capacity as function of modulation parameter $\tau$ for various noise $\SNReta=15$,25,35 and 50.}
\label{fig:memory}
\end{figure*}

Figures \ref{fig:memory}(a-d) show the memory capacities, i.e., $C_\mathrm{linear}$, $C_\mathrm{quad}$, $C_\mathrm{cross}$ and $C_\mathrm{total}$, of the PhRC-SBS system for various pump amplitudes, $\ahat_p$  ,and Stokes' amplitudes, $\ahat_s$. The total capacity is defined as $C_\mathrm{total}=C_\mathrm{linear}+C_\mathrm{quad}+C_\mathrm{cross}$, as in \cite{paquot2012optoelectronic,duport2012all}. Specifically for Figs. \ref{fig:memory}(a-d), the modulation parameter $\tau=160\gamma_a^{-1}$ and an effective $\mathrm{SNR}_\eta=35$ dB have been used.  Figures \ref{fig:memory}(a-d) show that in all cases, the optimum memory capacity occurs at the edge of the bifurcation line, shown as dashed lines, deviating by approximately 10\% due to the multiplexing parameter $\hat{\kappa}$. On each of Fig. \ref{fig:memory}(a-d), the optimum operation conditions within the parameters search space, $10^{-5}\leq\ahat_s\leq10^{-3}$ and $0.3\leq\ahat_p\leq0.5$, have been denoted by a star. 

\revision{Optimal operation of the SBS kernel for PhRC applications occurs at the edge of the bifurcation line, as indicated in Figs. \ref{fig:memory}(a-d). Additionally, the location of the bifurcation line depends on the length of the fibre; a shorter length results in the bifurcation line appearing at higher input laser intensities, while a longer length shifts it towards lower intensities, see Figs. \ref{fig:dynamic}(e-f). However, using longer fibres introduces extended signal delays, which increase the computational delay of the PhRC system. Therefore, for the practical realisation of the PhRC-SBS system, one must consider the damage threshold optical power of the fibre, acceptable computational delay, and power budgeting of the overall system. Alternatively, to enable SBS at lower injected power and shorter fibre, one can utilise specialty glass fibres with higher Brillouin gain \cite{hirano2009silica}.   } 

\begin{table}[htbp] 
\centering
\caption{\bf Maximum memory capacities for PhRC-SBS in comparison to other time-delayed based PhRC systems. For the PhRC-SBS capacity, $\SNReta=35$}
\begin{tabular}{p{0.15\linewidth}  p{0.22\linewidth}  p{0.25\linewidth}   p{0.19\linewidth} }
\hline
           & All-optical reservoir computing \cite{duport2012all} & Opto-electronic reservoir computing \cite{paquot2012optoelectronic} & The present work of PhRC-SBS \\
\hline
$\max C_\mathrm{linear}$ & 20.8	& 31.9 & 21.9 \\
$\max C_\mathrm{quad}$ & 4.16 & 4 & 3.1 \\
$\max C_\mathrm{cross}$ & 8.13 & 27.3 & 27.2 \\
$\max C_\mathrm{total}$ & 28.84 & 48.6 & 48.6 \\
\hline
\end{tabular}
  \label{tab:memory}
\end{table}

To investigate the impact of parameter $\tau$ and $\SNReta$ on the memory capacity, parameter sweeps as in Fig. \ref{fig:memory}(a-d) were repeated for different values of $\tau$ and $\SNReta$. The optimum memory capacity for each reconstruction task are plotted in Figs. \ref{fig:memory}(e-h). Figures \ref{fig:memory}(e-h) confirm that higher $\SNReta$ yields to a higher memory capacity in all cases. Generally, Figs. \ref{fig:memory}(e-h) show that low $\tau$ yields to a low memory capacity; increasing $\tau$ also increases the memory capacity but reaches a plateau at a different rate. For the linear, cross, and total capacities, further increases of $\tau$ leads to a gradual decrease in capacity, while for the quadratic memory, the capacity saturates. From Figs. \ref{fig:memory}(e-h), the maximum capacities for a moderate $\SNReta=35$ are listed in Table \ref{tab:memory}. To compare the capacities with other PhRC systems, Table \ref{tab:memory} also lists the maximum memory capacities from \cite{paquot2012optoelectronic,duport2012all} which are based on the time delayed architectures. The PhRC-SBS has a comparable memory capacity to other fibre based PhRC systems. 

\subsection{Prediction of NARMA system} \label{sec:narma}
Non-linear auto-regressive moving average (NARMA) systems are a class of non-linear dynamical system which have been extensively used as a model in the machine learning communities, including the PhRC, as a time-series prediction problem\cite{duport2016fully,paquot2012optoelectronic}. In this benchmark, the PhRC-SBS is trained to predict the future value of the tenth order NARMA model, i.e., NARMA-10, by, 
\begin{align} \label{eq:narma}
    \bar{y}(\tildet+1) = 0.3\bar{y}(\tildet) +0.05\bar{y}(\tildet) \left[ \sum_{k=0}^9 \bar{y}(\tildet-k) \right] +1.5\Sym(\tildet)\Sym(\tildet-9) +0.1 , \quad \tildet\geq 9
\end{align}
where the input information symbols are taken from a random number generator $\Sym\sim \mathcal{U}(0,0.5)$. The prediction task based on the NARMA-10 time series model (\ref{eq:narma}) is dependent on the past input signal $\Sym$ and past output values $y(\tildet-k)$ and so it is expected that an optimum operation condition to perform this task reflects the optimum operation condition in the memory reconstruction tasks. For the NARMA-10 benchmark, a stream of 3000 symbols $\Sym$ was generated from which the first 100 samples were discarded from the training to washout the transient, then the next 2500 samples were used for training and the last 400 samples were used for testing, as was done in \cite{duport2016fully,paquot2012optoelectronic}. The following parameters have been fixed and used throughout this task: number of virtual nodes $N_x=50$, and noise $\SNReta=35$. 

\begin{figure}[ht!]
\centering\includegraphics[width=0.65\textwidth]{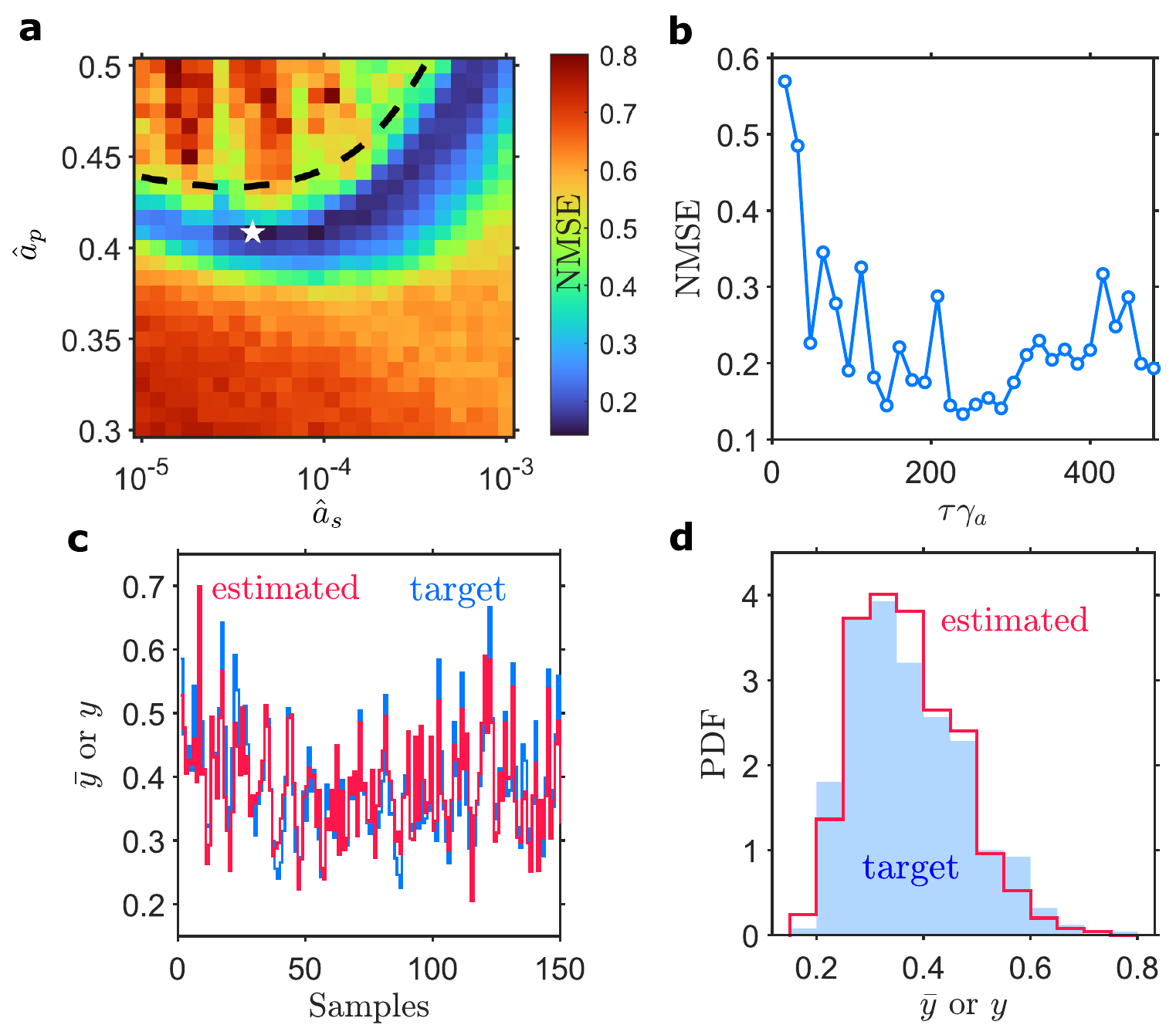}
\caption{PhRC-SBS performing NARMA-10. (a) The NMSE $\varepsilon$ as function of pump’s $\ahat_p$ and Stokes’ amplitude $\ahat_s$. The lowest error $\varepsilon$ is marked by a star. (b) The lowest $\varepsilon$ as function of modulation parameter $\tau$. (c) The first 150 samples of the predicted $y$ and target $\ybar$. (d) The probability distribution function (PDF) of the predicted $y$ and target $\ybar$ for all testing samples from (c). For (a,c,d), $\tau=144\gamma_a^{-1}$. For (c,d), $\ahat_p=0.41$ and $\ahat_s=4\times10^{-5}$.}
\label{fig:narma}
\end{figure}

Figure \ref{fig:narma}(a) displays the NMSE $\varepsilon$ of the PhRC-SBS performing NARMA-10 benchmark for different pumps amplitude $\ahat_p$ and Stokes' amplitude $\ahat_s$ for $\tau=144\gamma_a^{-1}$. It shows that the region of low error, i.e., high prediction rate, is located near the bifurcation line and has a similar shape to that found in the memory reconstruction tasks. Also, similar to the memory reconstruction task, there is an approximately 10\% deviation from the bifurcation line, which is due to the information modulation amplitude $\hat{\kappa}$. The optimum operation point is denoted by the star in the figure. To investigate the impact of the modulation parameter $\tau$ on the NARMA-10 task performance, Fig. \ref{fig:narma}(b) plots the minimum NMSE by repeating the parameter scan in Fig. \ref{fig:narma}(a) for various $\tau$ values. It shows that the error $\varepsilon$ decreases rapidly as $\tau$ increases and reaches a minimum $\varepsilon=0.13$ at $\tau=240\gamma_a^{-1}$. The PhRC-SBS has a higher prediction rate performance for the NARMA-10 benchmark compared to other previously reported PhRC configurations, for instance, $\varepsilon=0.168$ for the single delay line feedback loop with electro-optical nonlinear PhRC system\cite{paquot2012optoelectronic} and $\varepsilon=0.23$ for the all-optical fibre time-delayed PhRC\cite{duport2016fully}. To highlight the performance of the PhRC-SBS in this benchmark, Fig. \ref{fig:narma}(c) displays the desired target $\ybar$ and the predicted $y$ NARMA-10 time-series for the first 150 samples of the testing samples. The predicted values agree well with the desired series. Figure \ref{fig:narma}(d) shows the probability density function between the desired and predicted NARMA-10 output. It confirms that the PhRC-SBS can perform the NARMA-10 prediction with a good accuracy. 

\subsection{Non-linear channel equalisation task}

\begin{figure*}[ht]
\centering
\includegraphics[width=0.99\textwidth]{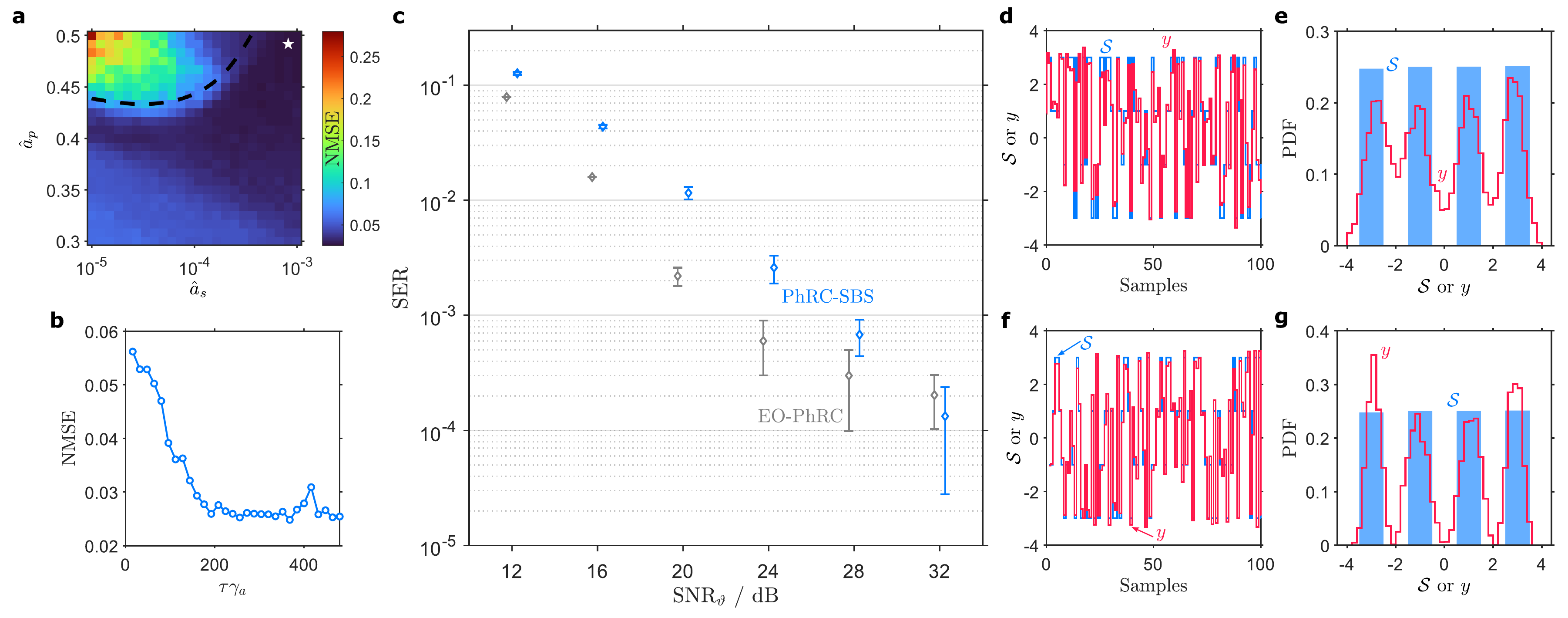}
\caption{Application of the PhRC-SBS for wireless non-linear channel equalisation task. (a) The NMSE $\varepsilon$ for both function of pump’s and Stokes’ amplitude $\ahat_{p,s}$, operated with $\tau=192\gamma_a^{-1}$ and $\SNReta=35$. The maximum capacity found within the parameter space scan has been marked by a star. (b) The lowest $\varepsilon$ as function of modulation parameter $\tau$. (c) The SER of the PhRC-SBS in comparison with electro-optics based EO-PhRC\cite{paquot2012optoelectronic} for various channel noise $\SNRtheta=12$,16, 20, 24, 28, and 32. Error bar shows the standard deviation of 10 independent train/test operations as in\cite{paquot2012optoelectronic}. (d,f) The first 100 samples of the predicted $y$ and target $\Sym$ for $\SNRtheta=16$ and 32, respectively. (e,g) The probability distribution function (PDF) of the predicted $y$ and target $\Sym$ for all testing sample for $\SNRtheta=16$ and 32, respectively. For comparison purposes (c-g) use $\eta=0$.}
\label{fig:nce}
\end{figure*}

For this benchmark task, the PhRC-SBS is used to perform a non-linear channel equalisation (NCE) task. NCE is an important signal processing task in wireless communication. It is used to recover information symbols which are distorted during its multipath propagation from transmitter to the receiver, i.e., multipath fading. Multipath fading is an important communication challenge and the PhRC can be used to address this issue at the edge of the data-infrastructure landscape. For this benchmark, multipath fading is modelled using the scheme defined in \cite{jaeger2004harnessing,paquot2012optoelectronic,jin2022adaptive} and is briefly described as follows: A sequence of information symbols is randomly generated from a set of possible values, i.e., $\Sym\in\{-3,-1,1,3\}$. Due to multipath signal arrival, the signal undergoes inter-symbol interference distortion, 
\begin{align}
    q(\tildet) &= 0.08 \Sym(\tildet+2) - 0.12\Sym(\tildet+1) + \Sym(\tildet) + 0.18\Sym(\tildet-1) \notag\\
               &-0.1\Sym(\tildet-2) +0.091\Sym(\tildet-3) -0.05\Sym(\tildet-4) \notag\\
               &+0.04 \Sym(\tildet-5) + 0.03\Sym(\tildet-6) + 0.01 \Sym(\tildet-7) ,\quad \tildet\geq 7.
\end{align}
It then goes through a noisy non-linear channel producing,
\begin{align}
    \Tilde{q}(\tildet) = q(\tildet) + 0.036q^2(\tildet) -0.011q^3(\tildet) +\vartheta,
\end{align}
where, $\vartheta$ is the communication channel noise and given as a zero-mean Gaussian distributed noise adjusted in power to yield to a particular signal-to-noise ratio of the communication channel ($\SNRtheta$). In this benchmark, the distorted information bits $\Tilde{q}$ is encoded to $u(t)$ following the multiplexing technique described in Section 2.\ref{sec:in} and modulated onto the pump laser. The PhRC-SBS is, then, trained to reconstruct the distorted information bits with the following target signal, 
\begin{align}
    \bar{y}(\tildet) = \Sym(\tildet).
\end{align}

To find the optimum PhRC-SBS operational parameters, as in the previous benchmark task, a stream of 3000 information symbols was generated from which the first 80\% were used for training and the remaining 20\% were used for testing. A similar procedure to that described in Section 4.\ref{sec:narma} was used to find the optimum operation parameters, i.e., scanning over the parameters space $\ahat_p$, $\ahat_s$ and $\tau$. The following parameters were fixed and used: number of virtual nodes $N_x=50$ and PhRC-SBS noise $\SNReta=35$. Figure \ref{fig:nce}(a) presents the NMSE as a function of pump amplitude $\ahat_p$ and Stokes amplitude $\ahat_s$ for a specific value of $\tau=192\gamma_a^{-1}$. It shows that  the suitable operation region for the NSE task also occurs around the bifurcation line. The optimum operation with least error $\varepsilon=0.025$ is marked by a star. To obtain the suitable information modulation parameter $\tau$, Fig. \ref{fig:nce}(b) plots the minimum NMSE $\varepsilon$ within the search parameter space $(\ahat_p,\ahat_s )$ as in Fig. \ref{fig:nce}(a) for various $\tau$ parameters. It shows that $\varepsilon$ rapidly decreases as $\tau$ increases and saturates at $\tau\approx192\gamma_a^{-1}$ beyond which the NMSE is in general $\varepsilon<0.03$.   

Comparison of the PhRC-SBS performance for NCE against other PhRC implementations is now made. From Fig. \ref{fig:nce}(b), the optimum operational parameters of $\ahat_p=0.49$ and $\ahat_s=8.1\times10^{-4}$ are used to perform NCE of a stream of 9000 symbols. Here, the first 3000 symbols are used for training and the remaining 6000 are used for testing as also in \cite{paquot2012optoelectronic}. The symbol error rate (SER), which is defined as the ratio of the total mistakenly reconstructed symbols $\Tilde{\Sym}$ by the total number of testing information symbols $\Sym$, is used as a quantitative measure for comparison purposes. To obtain the reconstructed symbols $\Tilde{\Sym}$ from the estimated signal $y$, an equidistant thresholding scheme was used, as also in\cite{jaeger2004harnessing,paquot2012optoelectronic}, that is the symbol ``1'' was chosen if $0<y<2$, symbol ``3'' was chosen if  $2<y<4$, etc.  The PhRC-SBS has a comparable performance for NCE applications to the EO-PhRC based on electro-optical system, as shown in Fig. \ref{fig:nce}(c). As the SER of the EO-PhRC is based on a first-principle model, with no presence of system noise $\SNReta$, the SER results for PhRC-SBS also calculated for $\eta=0$. Figures \ref{fig:nce}(d,e) and Figs.~\ref{fig:nce}(f,g) compare the performance the PhRC-SBS at two $\SNRtheta$ (16 and 32, respectively). They show that high $\SNRtheta$ leads to higher symbol reconstruction rate. The PDFs, in Fig. \ref{fig:nce}(e,g), show that in the lower $\SNRtheta$ case, there is a strong overlap in the distribution of the estimated signal y which leads to a lower reconstruction rate. In contrast, Fig. \ref{fig:nce}(g) shows that there are four distinct distributions which each correspond to the four target symbols $\Sym$ which yields an excellent symbol reconstruction. 

\revision{Based on the benchmarks conducted in Section \ref{sec:benchmark},  an important observation can now be made regarding the suitable information modulation parameter $\tau$. Since the information encoding approach employed in the new PhRC-SBS system uses a time-multiplexing scheme in time-delay-based PhRC systems \cite{larger2012photonic,paquot2012optoelectronic,appeltant2011information,anufriev2022non}, the benchmark results confirm that the suitable information modulation parameter $\tau$ for the PhRC-SBS system exhibits an intrinsic timescale linked to the signal propagation time within the fibre i.e. $\tau\gamma_a\approx L$.   } 

\begin{figure}[h!]
\centering\includegraphics[width=0.6\textwidth]{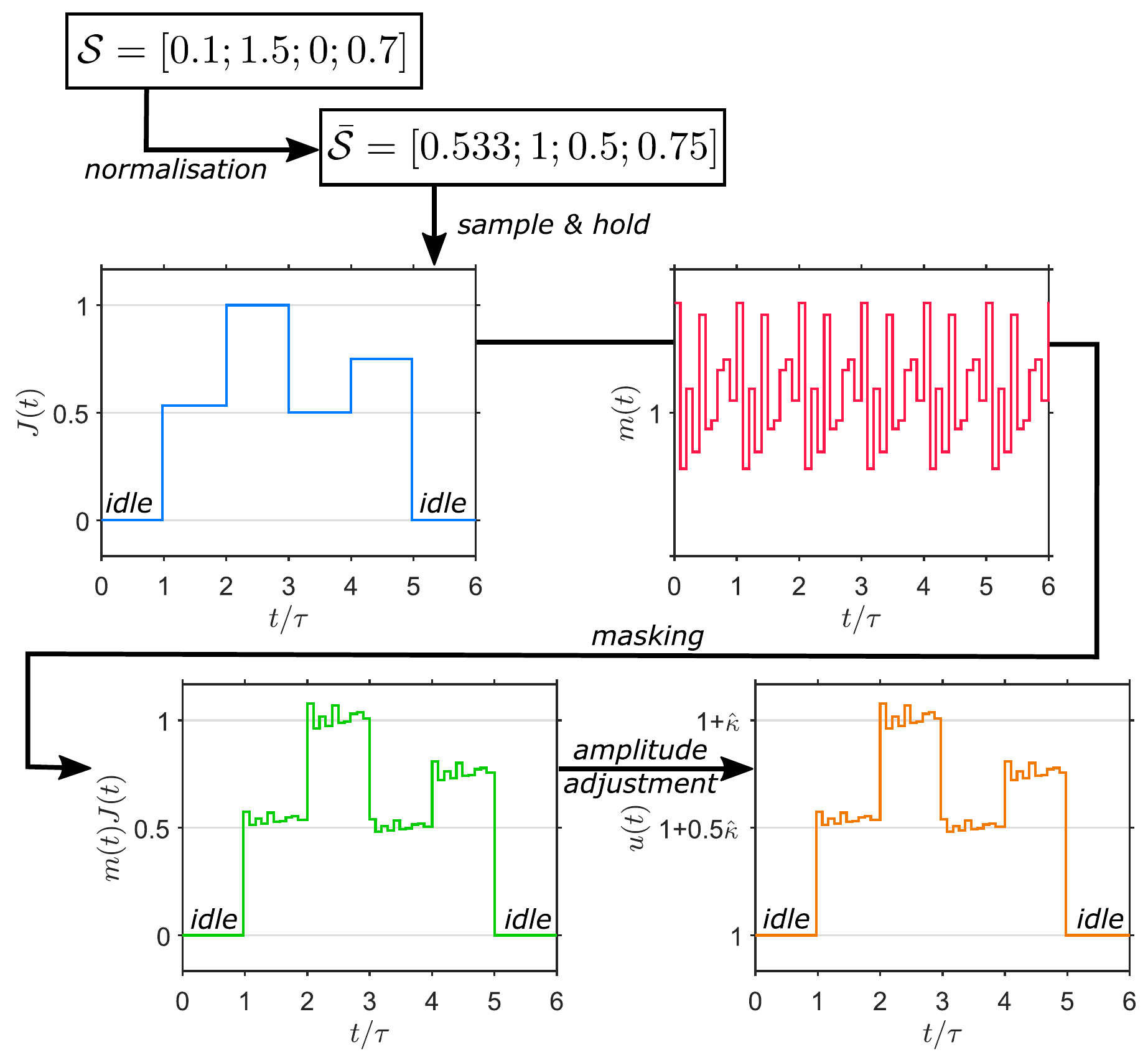}
\caption{Illustrative example of the optical multiplexing modulation procedure used in the present work.}
\label{fig:example}
\end{figure}

\section{Conclusions}
A new passive photonic reservoir computing architecture based on a stimulated Brillouin scattering system is described and numerically demonstrated. By solving the three-wave interaction model, the underlying property of the SBS system is analysed and shown to exhibit a phase transition from \revision{single} to \revision{multi-states} operation. It further shows the strong dependency of the PhRC-SBS’s performance to the phase transition of the SBS system, as such in all benchmark task the optimum operation condition is found to be at the edge of the bifurcation condition. It shows that the new PhRC-SBS system has a comparable or better performance in comparison to other reported PhRC systems.

\section*{Appendix: Time-multiplexing scheme}  \label{app:mul}

For example, let a four symbols information stream be given by $\Sym=[0.1;1.5;0;0.7]$. First a normalisation is performed via $\bar{\Sym}=(1-\chi)\frac{\Sym-\min[\Sym]}{\max[\Sym] - \min[\Sym]} + \chi$. Using $\chi=0.5$, as is used throughout the current work, leads to $\bar{\Sym}=[0.533;1;0.5;0.75]$. Then, the normalised symbols $\bar{\Sym}$ undergo a sample and hold operation which produces an analogue piecewise signal $J$ with width of $\tau$. A periodic mask $m$ is applied on to $J$ with $\hat{m}=0.1$, producing a small fluctuation within each $\tau$ duration. Finally, the masked signal is adjusted by $u(t)=1+\hat{\kappa} \hat{m}(t)J(t)$, with $\hat{\kappa}=0.1$, to allow for modulation using electro-optics modulator device. Figure \ref{fig:example} illustrates this procedure.

\begin{backmatter}
\bmsection{Funding} This work was supported by the Engineering and Physical Sciences Research Council (EP/V048937/1).

\bmsection{Acknowledgments} S.P. thanks TM Benson for proofreading the manuscript. 

\bmsection{Disclosures} The author declares no conflicts of interest.

\bmsection{Data Availability Statement} Data underlying the results presented in this paper are not publicly available at this time but may be obtained from the authors upon reasonable request.

\end{backmatter}

%%%%%%%%%%%%%%%%%%%%%%% References %%%%%%%%%%%%%%%%%%%%%%%%%

%%%%%%%%%% If using BibTeX:
\bibliography{referenceList}

%%%%%%%%%% If preparing manually:
% \begin{thebibliography}{1}
% \newcommand{\enquote}[1]{``#1''}

% \bibitem{Zhang:14}
% Y.~Zhang, S.~Qiao, L.~Sun, Q.~W. Shi, W.~Huang, L.~Li, and Z.~Yang,
%   \enquote{Photoinduced active terahertz metamaterials with nanostructured
%   vanadium dioxide film deposited by sol-gel method,}
%   {\protect\JournalTitle{Optics Express}} \textbf{22}, 11070--11078 (2014).

% \bibitem{Optica}
% {Optica}, \enquote{{Optica Publishing Group},}
%   \url{http://www.opg.optica.org}.

% \bibitem{FORSTER2007}
% P.~Forster, V.~Ramaswamy, P.~Artaxo, T.~Bernsten, R.~Betts, D.~Fahey,
%   J.~Haywood, J.~Lean, D.~Lowe, G.~Myhre, J.~Nganga, R.~Prinn, G.~Raga,
%   M.~Schulz, and R.~V. Dorland, \enquote{Changes in atmospheric consituents and
%   in radiative forcing,} in \enquote{Climate Change 2007: The Physical Science
%   Basis. Contribution of Working Group 1 to the Fourth assesment report of
%   Intergovernmental Panel on Climate Change,}  S.~Solomon, D.~Qin, M.~Manning,
%   Z.~Chen, M.~Marquis, K.~B. Averyt, M.~Tignor, and H.~L. Miler, eds.
%   (Cambridge University Press, 2007).

% \end{thebibliography}

\end{document}